\pdfoutput=1
\documentclass{article}

\usepackage{PRIMEarxiv}

\usepackage[utf8]{inputenc}
\usepackage[T1]{fontenc}
\usepackage[hidelinks]{hyperref}
\usepackage{url}
\usepackage{booktabs}
\usepackage{amsfonts}
\usepackage{amssymb}
\usepackage{amsmath}
\usepackage{nicefrac}
\usepackage{microtype}
\usepackage{graphicx}
\usepackage{float}
\usepackage{xcolor}
\usepackage[numbers,sort&compress]{natbib}
\usepackage{fancyhdr}
\usepackage{titlesec}
\usepackage{tikz}
\usepackage{tcolorbox}
\usetikzlibrary{arrows.meta,positioning,fit,backgrounds,calc,shapes.geometric}

\titlespacing*{\section}{0pt}{1.2ex}{0.8ex}

\newcommand{\figtt}[1]{\textup{\ttfamily #1}}

\title{Agentic Attribution of Cloud Resource Ownership}

\author{
  Tiltan Gilat, Gal Engelberg, Ariel Zadok, Leon Goldberg \\
  Sola Security \\
  \texttt{Corresponding author: gal.e@sola.security}
}

\hypersetup{
  pdftitle={Agentic Attribution of Cloud Resource Ownership},
  pdfauthor={Tiltan Gilat, Gal Engelberg, Ariel Zadok, Leon Goldberg},
  pdfkeywords={ownership attribution, security remediation, organizational context, cloud security, LLM agents, evaluation without ground truth}
}

\begin{document}
\maketitle

\begin{abstract}
Security tools raise findings across cloud environments, and AI agents now investigate and prioritize them. A finding is still resolved by people: it has to reach the people responsible for fixing it and accountable for the resource it concerns. Ownership is the organizational context that decides who that is. Where recorded, it is declared by hand in configuration management databases or owner tags and goes stale unless kept current; accounts are shared, and automation creates much of the infrastructure. The knowledge sits in work records, code history, audit logs and identity directories, encoded in each organization's own conventions. Recovering it poses four challenges: the signal is scattered across sources, none of them authoritative; the conventions that carry it differ between organizations; reaching an owner takes multi-hop links; and those related to a resource differ in the strength of their relation. We define ownership as a set of entities, people or teams, each related to the resource by one of five patterns ordered by strength. We present a reference architecture for agentic ownership attribution that derives ownership from these records instead of declaring it: it profiles an organization offline into tested signal procedures and an ownership playbook, then attributes each resource with a coordinator agent and parallel investigators, with confidence computed by rules. On 100 resources of 50 Amazon Web Services (AWS) resource types in one organization, we pool the architecture's candidates with those of an unprimed general-purpose agent and verify every evidence piece against the records. The architecture reaches 92.1\% weighted recall and 86.6\% weighted precision, against 54.8\% and 77.4\%, at 0.55 times the agent's cost per unit of coverage. The difference lies in people who build and change resources without an explicit owner, found through work records and code history.
\end{abstract}


\section{Introduction}
\label{sec:intro}

Security programs run on findings. Posture management, vulnerability scanners and detection tools raise them across cloud environments, each attached to a resource, and AI agents now investigate and prioritize them \citep{engelberg2026ani,goldberg2026coding}. A finding is still resolved by people: it has to reach the people responsible for fixing it and accountable for the resource, who decide whether an exposure is intended, have it fixed or accept the risk. Who they are is \emph{organizational context}, knowledge of what an organization's assets mean to it that changes the correct security decision: the same finding on two identically configured resources goes to different people, depending on who owns each \citep{yavin2026orgcontext}. Misallocation is costly. In a vendor survey of 300 cybersecurity and IT professionals, about half spend at least as much time coordinating remediation, including identifying owners, as analyzing risk \citep{seemplicity_2026report}, and more than 20{,}000 released cloud resources that remained referenced have been found hijacked \citep{friess2024cloudy}.

Yet ownership is rarely recorded reliably. Where it is recorded, it is declared: configuration management databases (CMDBs) and asset inventories hold owner fields \citep{servicenow_cmdbci} that people fill in by hand or copy from other declared records \citep{nucleus_teams}. The declaration ages as the most suitable owner changes over time \citep{ahlgren2020ownership}, and CMDB tooling has to measure how many records have gone stale \citep{servicenow_cmdbhealth}. Owner tags are missing or inconsistent; in the organization evaluated here, the standard owner and team tags are not in use at all. Accounts are shared, and automation creates much of the infrastructure. The knowledge sits instead in the records people leave as they work: the cloud audit log, the identity center, work tracking, code hosting and the identity provider. We call this \emph{implicit ownership}, ownership that these records hold but the resource does not; it is the ownership instance of the \emph{organizational context gap} \citep{yavin2026orgcontext}.

Recovering implicit ownership poses four challenges. The signal is scattered (C1): each source holds a partial, sometimes outdated trace, and none is authoritative. Conventions are organization-specific (C2): one organization encodes the owner in a tag, another in a resource name, a third only in tickets, so fixed rules do not transfer. Links are multi-hop (C3): a tag leads to a deployment stack, a repository, its pull requests, their authors and finally a current employee. Relations differ in strength (C4): the engineer who built a resource and a person who made a single commit to a library it uses should not weigh equally.

Existing work addresses neighboring questions. Software engineering measures code ownership from contribution shares \citep{bird2011dont,koana2024examining}, and triage assigns transient work items from historical assignments \citep{anvik2006who,chen2019empirical}. The closest academic precedent we found suggests owners for internal software assets at Facebook \citep{ahlgren2020ownership}. Products record or infer one owner per resource or finding, usually a team \citep{backstage_descriptor,datadog_ownershipagent,brinqa_attribution}, and none of those we reviewed publishes an evaluation method.

Formally, an organization integrates a set of sources $D$, and for a resource $r$ the task is to return its ownership set $A(r) = \{(e, E_e, c_e)\}$: each entity $e$, a person or a team, that records in $D$ relate to $r$, the evidence pieces $E_e$ that cite those records, and a confidence $c_e \in [0,1]$. The relation pattern $\pi(e,r)$, read from $E_e$, decides how much $e$ counts. Non-human identities, such as service accounts and roles used by software, and people no longer active in the organization are never members of $A(r)$.

This paper makes two contributions. First, we define ownership as a typed set of entities, people and teams: five relation patterns, ordered by strength, name the nature of each entity's tie to a resource. Second, we present a reference architecture for agentic ownership attribution that derives ownership from the records the organization already keeps, instead of a field that people must keep current. Running on a data layer that relates the organization's records across platforms, it profiles each organization offline into \emph{signal procedures}, tested lookups from a resource to a person or team, and an \emph{ownership playbook}, the guidance its agents read. It then attributes each resource with a coordinator agent and parallel investigators, while code checks each conclusion and computes confidence by rules.

The evaluation follows the pooling practice of information retrieval \citep{sparckjones1975report}. The candidates of the architecture and of an unprimed general-purpose agent are pooled. The agent is Claude Code \citep{anthropic_claudecode}, run through the Claude Agent SDK \citep{anthropic_agentsdk} on Claude Sonnet 5.5 \citep{anthropic_sonnet55}, the model of the architecture's inference agents, with direct read-only access to the organization's live platforms and no knowledge of the organization. In the pool, every evidence piece is verified against the records, and every verified owner is weighted by relation pattern. The evaluation covers 100 resources of 50 Amazon Web Services (AWS) resource types \citep{aws_resourcetypes} in one organization, and the discussion answers each challenge from its results.

Section~\ref{sec:related} reviews related work, Section~\ref{sec:definition} defines ownership, Section~\ref{sec:architecture} presents the architecture, Section~\ref{sec:evaluation} reports the evaluation, and Section~\ref{sec:discussion} discusses the results and their limits.

\section{Related Work}
\label{sec:related}

\paragraph{Organizational context for security agents.}
Our prior work develops the context that security agents reason over when they investigate and prioritize findings. An asset intelligence layer turns heterogeneous security data into asset-level abstractions such as blast radius and business and data criticality \citep{engelberg2026ani}. Benchmarks measure agents that investigate identity posture in a live identity environment \citep{engelberg2026solaispm} and across eight vendor platforms \citep{yavin2026crossvendor}, and an open framework benchmarks agents that investigate enterprise security posture \citep{engelberg2026osb}. On live cloud investigations, a context layer resolved offline covers more of a pooled answer set than general-purpose coding agents given read-only access to the same live AWS environment \citep{goldberg2026coding}. Organizational context is defined for security reasoning and extracted as claims from what people write (evaluated on policies, audit reports and architecture documents), carried with their evidence and bound to the assets they concern \citep{yavin2026orgcontext}. That work names ownership as organizational context and notes that part of it can be inferred from machine state. This paper defines ownership as a typed set of entities and recovers it, for each resource, from the organization's operational records.

\paragraph{Ownership in software engineering.}
Software engineering treats ownership mainly as a measurement: the proportion of a component's changes that one developer made. \citet{bird2011dont} relate such measures to failures in Windows Vista and Windows 7; replications confirm the relation in some settings \citep{greiler2015code} and find it weak in others \citep{foucault2014code}. The measured owner depends on the operationalization. Adding review activity shows that most developers who contribute to a module authored no code in it and only reviewed its changes \citep{thongtanunam2016revisiting}, and commit-based and line-based measures yield different owner sets \citep{thongtanunam2024code}. Declared ownership of code paths, through CODEOWNERS files \citep{github_codeowners,lulla2025codeowners} or OWNERS files kept current by activity rules \citep{chromium_owners}, assigns reviewers for code. These works define, measure or declare the ownership of code. The object here is a deployed resource, and the relation is established across systems that code-ownership measures do not read.

\paragraph{Expert finding and work-item triage.}
Expertise location answers who knows about a part of a system: from change-management data \citep{mockus2002expertise}, from a model of organizational expertise referral \citep{mcdonald2000expertise}, and from authorship combined with interaction in the development environment \citep{fritz2010degree}. Enterprise expert finding ranks people for a topic through person--document associations, whose strength is the central modeling choice \citep{balog2006formal,balog2012expertise}; reviewer recommendation \citep{thongtanunam2015who} and enterprise socio-technical graphs \citep{maddila2022nalanda} apply the same idea at scale. Bug triage maps a new report to a developer learned from past assignments \citep{cubranic2004automatic,anvik2006who} and models reassignment as tossing graphs \citep{jeong2009improving}. Incident triage routes incidents in online services to the responsible team \citep{chen2019empirical,pham2020deeptriage,gao2020scouts}; LLM agents now perform root-cause analysis that names responsibilities \citep{wang2024rcagent} and multi-agent incident triage \citep{yu2025triangle}. In all of these, the object is a topic or a transient work item and the output is the person or team to consult or to act next; standing ownership of a persistent resource has no assignment history to learn from.

\paragraph{Ownership of assets, services and data.}
\citet{ahlgren2020ownership} describe ownership management for internal software assets at Facebook, now Meta (source files, warehouse tables, configurations), where a deployed system mines internal records at scale with machine learning; they set out the open problems of the task. Configuration management databases, for which no common understanding of what constitutes one has been established \citep{brenner2006cmdb}, store ownership as declared attributes \citep{servicenow_cmdbci}, and their tooling scores stale or incomplete records rather than inferring the owner \citep{servicenow_cmdbhealth}. Dataset catalogs crawl owners as metadata and propagate them across related datasets \citep{halevy2016goods}, and data governance designs owner and steward roles as responsibility matrices \citep{weber2009one,khatri2010designing}. Measurement studies document the exposure created by resources that outlive their owners' attention \citep{liu2016dns,friess2024cloudy}. Our search found no peer-reviewed work that attributes deployed cloud resources to people.

\paragraph{Industry approaches.}
Products approach resource ownership through four mechanisms. Cloud providers govern declared metadata: tag policies standardize keys and values \citep{aws_tagpolicies}, tagging guidance defines separate owner tags for separate responsibilities \citep{aws_taggingwhitepaper}, and ownership labels are described as a common way to identify responsible teams \citep{gcp_labels}; remediation platforms resolve one owner from such declarations in a fixed order of trust, a manual override, possibly inherited from an account or network, before a CMDB record and owner tags \citep{tamnoon_ownership}. Creator provenance is recorded from control-plane events \citep{aws_createdby,azure_systemdata,custodian_autotaguser} as the principal that issued the create call, which is often a role session or a pipeline. Vendors describe code-to-cloud tracing that links resources derived from infrastructure-as-code and container images to repositories, commits and developers \citep{wiz_codetocloud,orca_cloudtodev,prisma_traceability,yor_github}. Owner inference goes further. Datadog's Ownership Agent \citep{datadog_ownershipagent} ranks evidence from owner tags, configured preferences, the service catalog, audit-log creators and naming, and returns one suggested owner with a confidence, an explanation and its evidence; organization-specific guidance is written by hand as a preferences table \citep{datadog_ownershippreferences}. Brinqa \citep{brinqa_attribution} predicts typed owner attributes with supervised models trained on existing assignments, Zafran \citep{zafran_ownership} describes an agent that cites its evidence and abstains when the evidence is insufficient, and developer portals \citep{cortex_owners,compass_owner} suggest owner teams from repository activity and linked work-tracking data.

Together, these lines of work place ownership in the organizational context that security judgments depend on, establish it as a consequential construct in software engineering, supply activity-based ways to rank people for an artifact, and show that owner inference is in commercial demand. What is missing is attribution with four properties: it treats the ownership of a deployed resource as a set of people and teams whose relations differ in kind and strength; it combines the organization's operational records for each resource; it learns where ownership signals live in a given organization instead of requiring them to be written by hand; and its accuracy is measured by a published method. These are the properties that \emph{implicit ownership} requires.

\section{Ownership Definition}
\label{sec:definition}

An owner of a resource is an entity, a person or a team, whose relation to the resource is supported by the organization's records. The definition makes ownership a set rather than a single value, and it makes the relation, not the entity, the unit that has to be justified. Each member of the set carries a \emph{relation pattern}, which names the nature of the entity's tie to the resource (Table~\ref{tab:patterns}).

Explicit owner and direct work form the strongest level. An explicit owner is named on the resource itself, in its name or recorded metadata, or is shown by a recorded action to have created it. Direct work covers the people who build, change, run or answer for the resource, or for the thing it exists for. The weaker patterns, in decreasing strength, are indirect work, the same activities on the resource's surroundings where the link is inferred; relies on it, use of or dependence on the resource; and peripheral, a weak link such as a mention.

The boundary between direct and indirect work separates the people who answer for a resource from those who answer for its surroundings. Work is direct when the thing the person or team works on is the reason the resource exists. Maintaining the database service that a compute node exists to run is direct work on the node; administering the cloud account in which the node runs is indirect work. A resource that exists for an environment, such as the environment's own certificate, is direct work for the person who runs that environment.

Five rules govern the patterns.
\begin{enumerate}
  \item\label{rule:behavior} Patterns name behaviors, not data sources, so a relation recorded in a source the definition did not anticipate still has a place.
  \item\label{rule:evidence} A pattern is read from the evidence that establishes the relation; assigning it requires no further investigation.
  \item\label{rule:level} Creating, changing, maintaining, operating and being responsible for a resource are one level; the builder is not ranked above the operator.
  \item\label{rule:breadth} Breadth of scope matters only relative to the other people related to the same resource. When nobody does direct work on a resource, indirect work is the strongest relation it has, and it is graded as such.
  \item\label{rule:nonhuman} Non-human identities are not owners. A deployment pipeline, an infrastructure-as-code role or a service principal points to the run it performed. The run points to a branch, commit or ticket, and the people on those records are the candidates.
\end{enumerate}

The order encodes a judgment about allocation. The strongest level holds the entities a finding on the resource should reach: those named as answering for it and those who build and run it. Missing one of them is a larger error than missing a person who once mentioned the resource. The evaluation turns this order into weights.

Existing ownership models characterize the owner: by artifact, owner and degree of ownership \citep{koana2024examining}, or by declared role, as owned by, managed by and supported by \citep{servicenow_cmdbci}, as one ultimate owner \citep{backstage_descriptor}, or as several owners of named types \citep{datadog_entitymodel,datahub_ownershiptypes}. The relation patterns instead type and grade the tie that the records establish between an entity and one resource, so the same entity can be the explicit owner of one resource and a peripheral link to another. Rule~\ref{rule:nonhuman} corresponds to delegation in provenance models, where an agent acts on behalf of another who retains some responsibility \citep{moreau2013provdm}.

\begin{table}[H]
\centering
\small
\caption{Relation patterns.}
\label{tab:patterns}
\begin{tabular}{@{}p{2.4cm}p{9.6cm}p{2.0cm}@{}}
\toprule
Pattern & Definition & Strength \\
\midrule
Explicit owner & The entity's name, username or email appears in the resource's name or in metadata recorded on it, or a recorded action shows the entity created it & Highest \\
Direct work & Builds, changes, runs or answers for this resource, or for the thing it exists for & Highest \\
Indirect work & Builds, changes, runs or answers for something around the resource (its environment, the infrastructure beneath it, related resources); the link to this resource is inferred & Intermediate \\
Relies on it & Uses the resource or its system, builds on top of it, or reports problems while using it & Low \\
Peripheral & A weak link (a mention, a look-alike), or no pattern fits & Lowest \\
\midrule
Non-human identity & Service account, machine user, role used by software & Not an owner \\
\bottomrule
\end{tabular}
\end{table}

\section{Reference Architecture}
\label{sec:architecture}

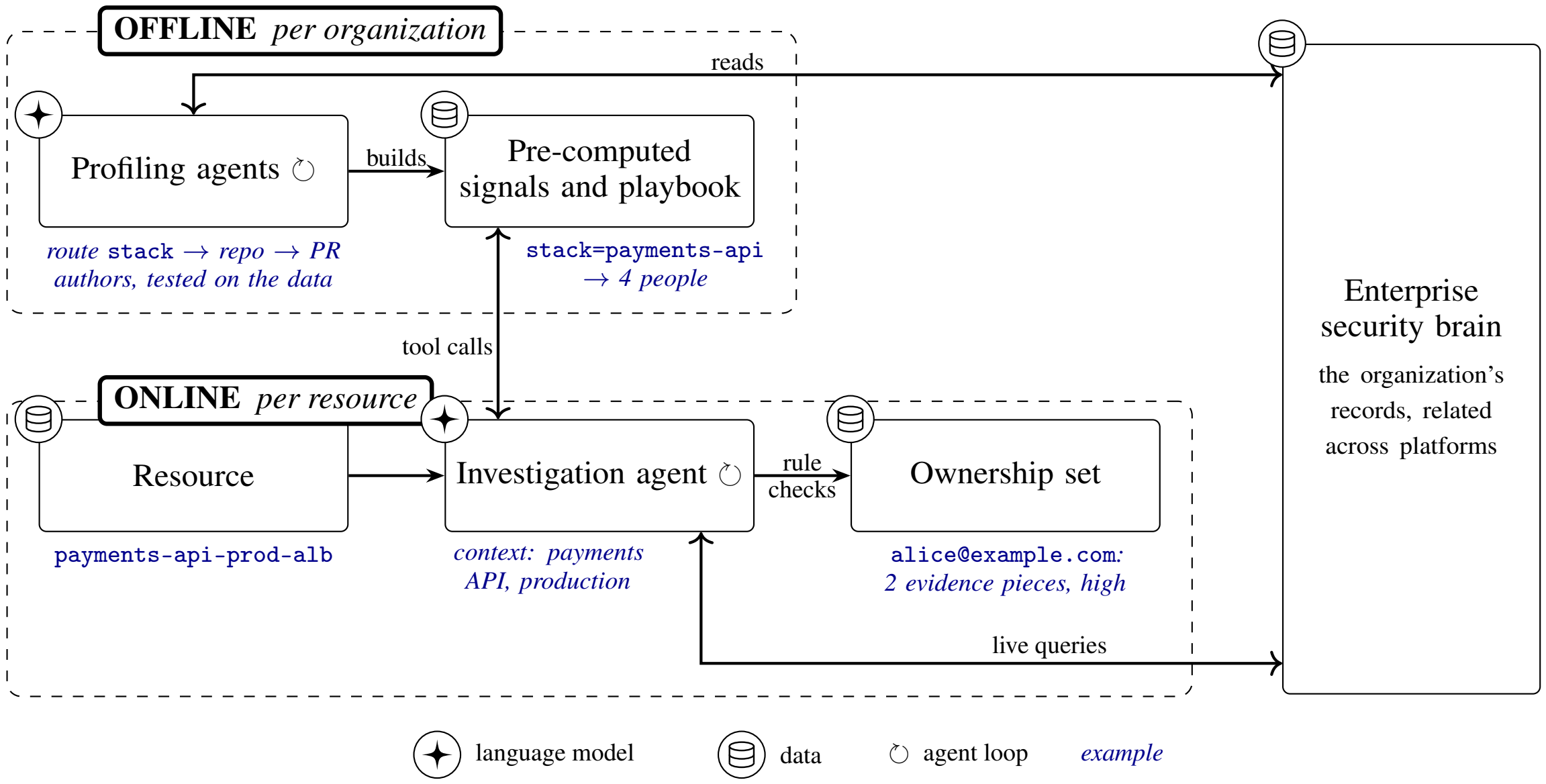
\begin{figure}[H]
\centering
\begin{tikzpicture}[
  font=\footnotesize,
  box/.style={draw, rounded corners=2pt, fill=white, align=center, minimum height=11mm, text width=2.9cm, inner sep=2pt, execute at begin node={\hyphenpenalty=10000}},
  ex/.style={font=\scriptsize\itshape, text=blue!55!black, align=center, inner sep=1pt, anchor=north},
  badge/.style={draw, circle, fill=white, minimum size=4.6mm, inner sep=0pt},
  iconllm/.pic={\fill[x=0.5mm,y=0.5mm] (0,3) .. controls (0.35,0.35) .. (3,0) .. controls (0.35,-0.35) .. (0,-3) .. controls (-0.35,-0.35) .. (-3,0) .. controls (-0.35,0.35) .. (0,3);},
  icondb/.pic={\begin{scope}[x=0.75mm, y=0.75mm, line width=0.45pt]
     \draw (0,1.1) ellipse (1.6 and 0.55);
     \draw (-1.6,1.1) -- (-1.6,-1.1); \draw (1.6,1.1) -- (1.6,-1.1);
     \draw (-1.6,-1.1) arc (180:360:1.6 and 0.55);
     \draw (-1.6,0) arc (180:360:1.6 and 0.55);
   \end{scope}},
  arr/.style={-{Stealth[length=1.8mm]}, thick},
  phase/.style={draw, very thick, rounded corners=2pt, fill=white, anchor=west, font=\small, inner xsep=4pt, inner ysep=2.5pt},
  note/.style={font=\scriptsize, align=center, inner sep=1pt}
]
\node[box] (prof) at (0,0) {Profiling agents $\circlearrowright$};
\node[box] (kno) at (4.0,0) {Pre-computed signals and playbook};
\node[ex, text width=3.1cm] (eprof) at ([yshift=-1mm]prof.south) {route \figtt{stack} $\to$ repo $\to$ PR authors, tested on the data};
\node[ex, text width=2.4cm] (ekno) at ([xshift=4.5mm, yshift=-1mm]kno.south) {\figtt{stack=payments-api}\\$\to$ 4 people};
\node[box] (res) at (0,-3.0) {Resource};
\node[box] (agt) at (4.0,-3.0) {Investigation agent $\circlearrowright$};
\node[box] (out) at (8.0,-3.0) {Ownership set};
\node[ex, text width=3.1cm] (eres) at ([yshift=-1mm]res.south) {\figtt{payments-api-prod-alb}};
\node[ex, text width=2.2cm] (eagt) at ([xshift=-5mm, yshift=-1mm]agt.south) {context: payments API, production};
\node[ex, text width=3.1cm] (eout) at ([yshift=-1mm]out.south) {\figtt{alice@example.com}:\\2 evidence pieces, high};
\node[draw, rounded corners=2pt, fill=white, align=center, text width=2.3cm, minimum height=6.4cm] (brain) at (12.0,-1.95) {Enterprise security brain\\[3pt]{\scriptsize the organization's records, related across platforms}};
\draw[arr] (prof) -- node[note, above] {builds} (kno);
\draw[arr, <->] ([xshift=-10mm]kno.south) -- node[note, left, pos=0.62] {tool calls} ([xshift=-10mm]agt.north);
\draw[arr] (res) -- (agt);
\draw[arr] (agt) -- node[note, above] {rule} node[note, below] {checks} (out);
\draw[arr, <->] (prof.north) |- node[note, above, pos=0.75] {reads} (brain.west |- 0,0.95);
\draw[arr, <->] ([xshift=10mm]agt.south) |- node[note, above, pos=0.8] {live queries} (brain.west |- 0,-4.85);
\coordinate (offtop) at (0,1.2);
\coordinate (onbot) at (0,-5.0);
\begin{scope}[on background layer]
\node[draw, dashed, rounded corners=4pt, inner xsep=7pt, inner ysep=5pt, fit=(prof)(kno)(eprof)(ekno)(offtop)] (off) {};
\node[draw, dashed, rounded corners=4pt, inner xsep=7pt, inner ysep=5pt, fit=(res)(agt)(out)(eres)(eagt)(eout)(onbot)] (on) {};
\end{scope}
\node[phase] at ([xshift=9mm]off.north west) {\textbf{OFFLINE}\enspace\textit{per organization}};
\node[phase] at ([xshift=9mm]on.north west) {\textbf{ONLINE}\enspace\textit{per resource}};
\foreach \n/\i in {prof/iconllm, kno/icondb, res/icondb, agt/iconllm, out/icondb, brain/icondb}
  {\node[badge] at (\n.north west) {}; \pic at (\n.north west) {\i};}
\node[badge] (l1) at (2.4,-5.75) {}; \pic at (l1) {iconllm}; \node[anchor=west, note] at ([xshift=1mm]l1.east) {language model};
\node[badge] (l3) at (5.4,-5.75) {}; \pic at (l3) {icondb}; \node[anchor=west, note] at ([xshift=1mm]l3.east) {data};
\node[anchor=west, note] at (6.8,-5.75) {$\circlearrowright$\; agent loop};
\node[anchor=west, note, text=blue!55!black] at (8.7,-5.75) {\itshape example};
\end{tikzpicture}
\caption{Agentic reference architecture: an investigation agent and its tools, one built offline, with an example at each step.}
\label{fig:architecture}
\end{figure}

The architecture treats ownership attribution as an investigation over the organization's records, carried out by an agent whose action space is a set of tools (Figure~\ref{fig:architecture}). Three principles shape it. First, judgment and guarantees are separated: language models decide where to look and how to read what they find, while deterministic code tests their proposals, decides whether a conclusion is admitted and combines evidence scores into a confidence. Second, organization-specific knowledge is compiled once: what can be learned about an organization's conventions is turned offline into tools, so that per-resource reasoning is spent only on what cannot be pre-computed. Third, every claim is grounded: each owner in the output carries the records that relate them to the resource, with facts kept apart from inferences.

\paragraph{Enterprise security brain.}
All agents and tools read the \emph{enterprise security brain}, the cross-platform security data layer of our prior work \citep{engelberg2026ani,goldberg2026coding,yavin2026orgcontext}. It integrates each platform's records in one read-only store, relates them in an asset graph before any question arrives, so that an identity-provider account already points to the cloud role it assumes, and holds organizational context beside the graph \citep{yavin2026orgcontext}. Relational context of this kind raised agent answer correctness by about 34\% relative in our cross-vendor benchmark \citep{yavin2026crossvendor}, and it lets questions over a whole population be answered against the complete inventory rather than a sample \citep{goldberg2026coding}.

\paragraph{Offline: compiling the organization into tools.}
Profiling agents learn how one organization records ownership (C2) and compile it into two tools. The first is a set of \emph{signal procedures}, executable lookups from a resource to a person or team. Each procedure starts as a hypothesis that an agent proposes about the organization, for example that the value of a \texttt{stack} tag names a repository whose pull-request authors work on the resource, and becomes a tool only after code, running it on the organization's own data, finds that it reaches people, generalizes beyond the values it was built from and behaves as declared. A procedure is defined over keys and joins, not over particular values, so one procedure serves every resource that carries the key. A hypothesis the data refutes is revised or dismissed. Accepted procedures are executed for every resource ahead of time as \emph{pre-computed signals}, so the deterministic part of attribution is paid once, and each carries a reliability prior derived from its known weaknesses. Signals differ in what they assert: a \emph{direct} signal names one person, as an owner tag holding \texttt{alice@example.com} does; a \emph{supportive} signal points toward owners, as a stack tag does; and \emph{chain segments} compose into multi-hop chains (C3). The second tool is the \emph{ownership playbook}, the declarative complement to the procedures: written guidance on which tags are reliable, how logins map to people and which leads are dead ends, which turns profiling from a one-off analysis into context read for every resource.

\paragraph{Online: an agentic search for evidence.}
Inference for one resource is a search over the evidence the brain can produce, run by one investigation agent in a loop of reasoning, tool calls and observation \citep{yao2023react}. The agent first infers what the resource belongs to, its \emph{context}: a load balancer named \texttt{payments-api-prod-alb} belongs to the payments API in production, so \texttt{payments} and \texttt{api} become search terms for tickets and code (C1, C3). It reads the compiled tools and uses live read-only queries for what they do not cover, such as tickets that mention the payments API. Open leads are pursued in parallel by investigator sub-agents that run the same loop, in the orchestrator--worker pattern of recent multi-agent systems \citep{fourney2024magentic,anthropic2025multiagent}. Two behaviors keep the search broad rather than greedy: every role holder on a matching record (assignee, reporter, author, reviewer) is registered, not only its leader, and automation is read as delegation, so a pipeline that created the resource is followed to the commit or ticket behind it (rule~\ref{rule:nonhuman}). Every \emph{evidence piece} separates what the records show from what is inferred from them: \emph{the person is the assignee of a ticket that names the stack} is a fact, \emph{the person therefore works on the resource} an inference.

\paragraph{Invariants and confidence.}
Acceptance is governed by invariants enforced in code rather than by the model's judgment. A conclusion is admitted only if it rests on evidence the agent has observed, every failed lookup was retried or explained, and every person it mentions is attributed or ruled out; otherwise it is returned for another round. Confidence is never stated by the model: investigators score each evidence piece, and fixed rules combine these scores and the procedures' priors into a confidence per entity. The output is the ownership set $A(r)$: each entity, its evidence and its confidence, from which the relation pattern is read (rule~\ref{rule:evidence}).

\section{Evaluation}
\label{sec:evaluation}

We evaluate the ownership set the architecture returns for each resource against the set returned by an unprimed baseline agent. No ground-truth ownership map exists, so both are scored against a pool of their verified candidates. Three criteria apply, each with its metrics: \emph{coverage}, whether a system finds the entities the records relate to the resource, measured by weighted recall and by success rate, the share of resources with a verified owner; \emph{correctness}, whether what it proposes is supported by the records, measured by weighted precision and rejected candidates; and \emph{cost}, measured by cost per resource and per unit of weighted recall.

For a resource, let $V$ be its verified candidates, $S$ the candidates proposed by system $s$, $R_s$ the rejected candidates of $s$, $w(c)$ the weight of the relation pattern of candidate $c$, and $H(\cdot)$ the restriction to people. The primary metrics are
\begin{equation}
\mathrm{WR}_s = \frac{\sum_{c \in H(S \cap V)} w(c)}{\sum_{c \in H(V)} w(c)}, \qquad
\mathrm{WP}_s = \frac{\sum_{c \in H(S \cap V)} w(c)}{|H(S \cap V)| + |R_s|}.
\end{equation}
Weighted recall is the share of the pool's verified weight a system found; in weighted precision each verified person earns its weight and each rejected candidate costs a full 1. The weights turn the strength order of Section~\ref{sec:definition} into numbers, in the manner of graded relevance \citep{jarvelin2002cumulated}: explicit owner and direct work 1.0, indirect work 0.5, relies on it 0.25, peripheral 0.1. When a resource has no direct worker, indirect work counts 1.0 (rule~\ref{rule:breadth}); when it has a verified explicit owner, direct work counts 0.75, indirect work 0.25 and weaker links 0. Values are micro-averaged, so each owner counts once, and only people are scored.

\subsection{Experiment design}
\label{sec:design}

\paragraph{Resources.}
Resources are drawn from a real-life production environment of one organization. Of the 244 AWS resource types we classified, 105 are \emph{ownable}, meaning a person or team would create, deploy, manage or answer for their instances; inherited records, platform metadata, findings and platform-wide configuration are excluded. Each resource is assigned a \emph{signal level} by how many of its applicable signal procedures hold a value for it: rich (at least half), thin (fewer), or metadata only and opaque (none). A stratified selection draws 100 resources of 50 ownable types, two per type, spread over signal levels, avoiding near-duplicate copies of one stack and confirming that each still exists.

\paragraph{Data sources.}
Six sources hold the records that relate people to resources, and they play four roles. \emph{Resource state}: the AWS resource inventory describes each resource by its name, tags, configuration and relations to other resources (59,309 resources of ownable types in one account and 18 regions); these descriptors are where an investigation starts. \emph{Activity}: the AWS audit log records who created, changed or tagged a resource, over the last 90 days. \emph{Work}: Jira records who reported, planned and carried out work through ticket roles and text (16,590 tickets in 18 projects over two and a half years), and GitHub records who wrote and reviewed code through pull requests and commits (10,005 pull requests in 44 repositories over two years). \emph{Identity}: Okta, the identity provider, holds the canonical directory of people with their employment status (73 people), and the AWS identity center (75 people and the roles they act through) and GitHub membership (51 members) link cloud principals and code-hosting logins to these people. Both systems access them read-only.

\paragraph{Systems.}
The architecture reads the enterprise security brain, into which the sources are ingested and related ahead of time. Its investigation and procedure-construction agents run on Claude Sonnet 5.5 \citep{anthropic_sonnet55}, and tag triage and playbook writing on Claude Sonnet 4.6 \citep{anthropic_sonnet46}. The baseline agent is Claude Code \citep{anthropic_claudecode} run through the Claude Agent SDK \citep{anthropic_agentsdk} on Claude Sonnet 5.5, given only the resource identifier and read-only access to the live platforms through the AWS and GitHub command-line interfaces (CLIs), the Jira API and a monitoring API, within 30 turns and 20 minutes per resource. It receives no ownership definition, strategy or knowledge of the organization, so it measures what an unprimed agent achieves with direct access, as in our earlier comparison with coding agents \citep{goldberg2026coding}.

\paragraph{Pool construction.}
The candidates of both systems for a resource are merged into one pool by canonical email, following the pooling practice of information retrieval \citep{sparckjones1975report,harman1993overview}. Because the baseline agent contributes candidates the architecture did not find, the pool does not rest on the system under test alone. Recall against a pool is relative recall \citep{clarke1997estimating}: it ranks systems reliably and bounds true recall from above, since owners no system found remain invisible \citep{zobel1998reliable}.

\paragraph{Evaluation pipeline.}
The pool passes four stages (Figure~\ref{fig:pipeline}). People who have left the organization or are unknown to its directories are rejected. A judge on Claude Sonnet 5.5 then verifies each evidence piece on its own: it sees the cited records retrieved afresh, may consult the integrated copy and the live platforms alike, and decomposes the claim, as in atomic-fact evaluation \citep{min2023factscore,wei2024longform}, into claimed facts, each confirmed, contradicted or not shown by a verbatim excerpt, and inferences, which are listed but never judged. Absence of data is never a contradiction, and the judge never sees either system's confidence. A candidate is \emph{verified} when at least one piece is confirmed and none contradicted, \emph{rejected} when the reverse holds, and \emph{insufficient} otherwise. Each verified person is then assigned a relation pattern by Claude Sonnet 4.6, with three votes combined per person, and weighted. The judge matched all 9 calibration cases in each of 2 runs and all 23 hand-labeled borderline candidates; no agreement study with human annotators was made. Example~1 follows three resources through the pipeline.

\begin{figure}[H]
\centering
\begin{tikzpicture}[
  font=\footnotesize,
  box/.style={draw, rounded corners=2pt, align=center, minimum height=10mm, text width=2.35cm, fill=white},
  store/.style={draw, align=center, minimum height=7mm, text width=2.35cm, fill=gray!12},
  arr/.style={-{Stealth[length=1.8mm]}, semithick}
]
\node[store] (sys) at (0,0.55) {Architecture};
\node[store] (ref) at (0,-0.55) {Baseline agent};
\node[box] (pool) at (3.2,0) {Pool candidates by identity};
\node[box] (dir) at (6.4,0) {Reject people outside the organization};
\node[box] (judge) at (9.6,0) {Verify each evidence piece against the records};
\node[box] (cls) at (12.8,0) {Classify relation and weight};
\node[box] (score) at (12.8,-1.7) {Score each system against the pool};
\draw[arr] (sys.east) -- (pool.west);
\draw[arr] (ref.east) -- (pool.west);
\draw[arr] (pool) -- (dir);
\draw[arr] (dir) -- (judge);
\draw[arr] (judge) -- (cls);
\draw[arr] (cls) -- (score);
\end{tikzpicture}
\caption{Evaluation pipeline.}
\label{fig:pipeline}
\end{figure}
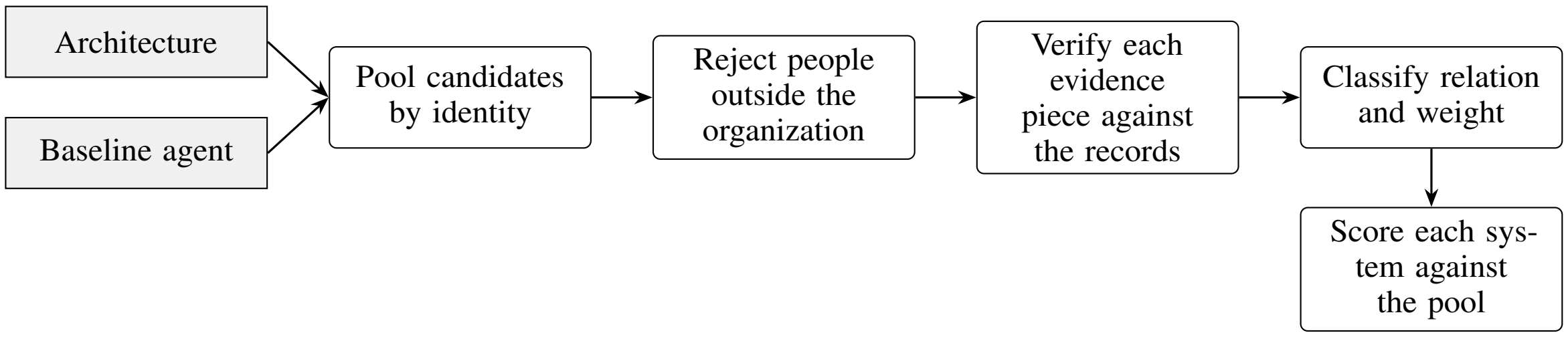

\begin{figure}[tp]
\begin{tcolorbox}[colback=white, colframe=black, boxrule=0.6pt, arc=2pt, left=3pt, right=3pt, top=3pt, bottom=3pt, fonttitle=\bfseries\small, title={Example~1: three resources, their verified owners and the weights that score them}]
\centering
\begin{tikzpicture}[
  font=\footnotesize,
  res/.style={draw, thick, rounded corners=2pt, align=center, text width=2.3cm, fill=gray!8, inner sep=3pt},
  ent/.style={draw, thick, rounded corners=2pt, align=center, text width=2.0cm, minimum height=6.6mm, inner sep=1.5pt},
  hd/.style={font=\scriptsize\bfseries, anchor=base, text depth=0pt, align=center},
  arr/.style={-{Stealth[length=1.5mm]}, semithick},
  ok/.style={font=\normalsize},
  no/.style={font=\normalsize, text=gray},
  wlab/.style={font=\scriptsize, anchor=west, inner sep=1pt}
]
\def\wx{4.65}\def\wu{1.6}
\node[hd] at (0,0.62) {Resource};
\node[hd] at (3.05,0.62) {Verified owner};
\node[hd] at (5.6,0.62) {Weight};
\node[hd] at (8.4,1.1) {Proposed by};
\draw[gray] (6.85,1.0) -- (9.95,1.0);
\node[hd] at (7.5,0.62) {Architecture};
\node[hd] at (9.3,0.62) {Baseline agent};
\node[hd] at (12.15,1.1) {Weighted recall};
\draw[gray] (10.45,1.0) -- (13.85,1.0);
\node[hd] at (11.2,0.62) {Architecture};
\node[hd] at (13.1,0.62) {Baseline agent};
\draw[gray!60] (-1.3,0.45) -- (13.85,0.45);
\foreach \n/\rel/\y/\w/\pre/\a/\b in {
  A/direct work/0/1.0/0/1/1, B/direct work/-0.8/1.0/0/1/0, C/direct work/-1.6/1.0/0/1/0,
  D/direct work/-2.75/1.0/0/1/0, E/indirect work/-3.55/0.5/0/0/1,
  F/explicit owner/-4.7/1.0/0/1/1, G/direct work/-5.5/0.75/1/1/0} {
  \node[ent] (p\n) at (3.05,\y) {\textbf{Person \n}\\[-1pt]{\scriptsize \rel}};
  \fill[black!60] (\wx,\y-0.13) rectangle (\wx+\w*\wu,\y+0.13);
  \ifnum\pre=1 \draw[black!60, dash pattern=on 1.2pt off 1pt] (\wx+\w*\wu,\y-0.13) -- (\wx+\wu,\y-0.13) -- (\wx+\wu,\y+0.13) -- (\wx+\w*\wu,\y+0.13); \fi
  \node[wlab] at (\wx+\wu+0.08,\y) {\w};
  \ifnum\a=1 \node[ok] at (7.5,\y) {$\checkmark$}; \else \node[no] at (7.5,\y) {$\times$}; \fi
  \ifnum\b=1 \node[ok] at (9.3,\y) {$\checkmark$}; \else \node[no] at (9.3,\y) {$\times$}; \fi
}
\node[res, minimum height=2.15cm] (r1) at (0,-0.8) {\textbf{VPN gateway}\\[2pt]{\scriptsize shared instance}\\{\scriptsize tag: deployment stack}\\{\scriptsize owner recorded: none}};
\node[res, minimum height=1.35cm] (r2) at (0,-3.15) {\textbf{Wildcard certificate}\\[2pt]{\scriptsize shared}\\{\scriptsize owner recorded: none}};
\node[res, minimum height=1.35cm] (r3) at (0,-5.1) {\textbf{Serverless function}\\[2pt]{\scriptsize name carries a person's handle}};
\foreach \r/\n in {r1/A, r1/B, r1/C, r2/D, r2/E, r3/F, r3/G} \draw[arr] (\r.east |- p\n) -- (p\n.west);
\draw[gray!60] (-1.3,-2.175) -- (13.85,-2.175);
\draw[gray!60] (-1.3,-4.125) -- (13.85,-4.125);
\foreach \y/\fa/\pa/\ra/\fb/\rb in {-0.8/3.0/3.0/100/1.0/33, -3.15/1.0/1.5/67/0.5/33, -5.1/1.75/1.75/100/1.0/57} {
  \draw[black!40] (10.45,\y+0.02) rectangle (11.95,\y+0.26);
  \fill[black!70] (10.45,\y+0.02) rectangle (10.45+\ra*0.015,\y+0.26);
  \node[font=\scriptsize, anchor=north] at (11.2,\y-0.04) {\textbf{\ra\%}\enspace{\tiny \fa/\pa}};
  \draw[black!40] (12.35,\y+0.02) rectangle (13.85,\y+0.26);
  \fill[black!30] (12.35,\y+0.02) rectangle (12.35+\rb*0.015,\y+0.26);
  \node[font=\scriptsize, anchor=north] at (13.1,\y-0.04) {\rb\%\enspace{\tiny \fb/\pa}};
}
\end{tikzpicture}

\smallskip
\raggedright\small
The gateway records no owner: all three people are reached through tickets and pull requests, Person~C only as the reporter of the right-sizing ticket and its parent epic, a role the investigators register on every matching record. On the certificate each system finds one person, and the weights separate them: the person who works on it counts twice the person who works around it. The function's name gives both systems its explicit owner; only the architecture also finds the colleague in direct work, whose weight drops from 1.0 to 0.75 (dashed) because an explicit owner exists.
\end{tcolorbox}
\end{figure}

\subsection{Results}
\label{sec:results}

Table~\ref{tab:headline} summarizes the results by criterion. The architecture leads on coverage and weighted precision and costs less per unit of coverage; the paragraphs below take the criteria in turn, and coverage is broken down by relation, source and kind of resource.

\paragraph{Coverage.}
The architecture reaches 92.1\% weighted recall against 54.8\% for the baseline agent (93.3\% against 68.5\% macro-averaged), and proposes a verified owner for 99 of 100 resources against 89. Of the 230 verified person--resource pairs in the pool, 98 were found by both systems, 100 by the architecture only and 32 by the baseline agent only.

\emph{By relation} (Table~\ref{tab:relation}). Both systems find the person named on the resource, 47 and 48 of 49 explicit owners. The difference is direct work: the architecture finds 131 of 141 people against 53, and 117 of 123 against 47 on resources without an explicit owner.

\emph{By source} (Table~\ref{tab:sources}). The added owners are found beyond the resource. Code history backs 46\% of the architecture's verified owners against 2\%, person directories 68\% against 17\%, and tickets 65\% against 52\%. The baseline agent reads GitHub through its CLI, yet none of its verified owners rests on its own lookups there; its few code-history links are pull requests named in Jira tickets.

\emph{Owners found only by the baseline agent.} Of these 32 people, 20 are weaker links (indirect work, relies on it or peripheral), and the most common routes to them are free-text ticket search (14 people) and neighboring resources (11). The architecture reached 11 of the 32 during its investigation without proposing them and never reached the other 21.

\emph{By kind of resource} (Table~\ref{tab:breakdown}). Across resource categories, the architecture's weighted recall stays between 88\% and 95\%, against 44\% to 72\% for the baseline agent. It does not fall with less pre-computed signal: 90\% on rich resources, 96\% on thin ones and 94\% where no procedure applies. Categories hold 13 to 21 resources, and no significance is claimed.

\paragraph{Correctness.}
Weighted precision is 86.6\% against 77.4\% (89.5\% against 81.3\% macro-averaged). Few proposals of either system are rejected, 12 against 10, of which 8 each by the directory check, so the baseline agent's lower weighted precision reflects fewer and weaker owners rather than more errors. The judge decided 250 of the 265 pooled candidates (94\%).

\paragraph{Cost.}
Costs are computed from the recorded token counts at uncached list prices \citep{anthropic_pricing} and cover per-resource attribution; profiling, run once per organization, is not included. A resource costs \$0.63 with the architecture (about 190K input and 25K output tokens) and \$0.68 with the baseline agent (about 320K input and 4K output tokens). Per unit of weighted recall, the cost is \$0.68 against \$1.24, so the architecture delivers coverage at 0.55 times the baseline agent's cost.

\begin{table}[H]
\centering
\small
\caption{Results by criterion (micro-averaged, 99 scored resources).}
\label{tab:headline}
\begin{tabular}{@{}llrr@{}}
\toprule
Criterion & Metric & Architecture & Baseline agent \\
\midrule
Coverage & Weighted recall & \textbf{92.1\%} & 54.8\% \\
 & Success rate & \textbf{99 of 100} & 89 of 100 \\
Correctness & Weighted precision & \textbf{86.6\%} & 77.4\% \\
 & Rejected candidates & 12 & 10 \\
Cost & Per resource, uncached list prices & \$0.63 & \$0.68 \\
 & Per resource, per unit of weighted recall & \textbf{\$0.68} & \$1.24 \\
\bottomrule
\end{tabular}
\end{table}

\begin{table}[H]
\centering
\small
\caption{Verified owners by relation pattern.}
\label{tab:relation}
\begin{tabular}{@{}lrrr@{}}
\toprule
Relation & Pool & Architecture & Baseline agent \\
\midrule
Explicit owner & 49 & 47 & 48 \\
Direct work & 141 & 131 & 53 \\
\quad on resources without an explicit owner & 123 & 117 & 47 \\
Indirect work, relies on it, peripheral & 40 & 20 & 29 \\
\midrule
Total & 230 & 198 & 130 \\
\bottomrule
\end{tabular}
\end{table}

\begin{table}[H]
\centering
\small
\caption{Share of verified owners backed by confirmed evidence from each source.}
\label{tab:sources}
\begin{tabular}{@{}lrr@{}}
\toprule
Source & Architecture (198) & Baseline agent (130) \\
\midrule
Cloud resource metadata: tags, names, linked resources & 81\% & 75\% \\
Person directories & 68\% & 17\% \\
Work-tracking tickets & 65\% & 52\% \\
Pull requests and commits & 46\% & 2\% \\
Code-hosting logins, teams and repositories & 28\% & 3\% \\
Cloud audit log activity & 18\% & 29\% \\
\bottomrule
\end{tabular}
\end{table}

\begin{table}[H]
\centering
\small
\caption{Weighted recall by resource category and signal level (one network, rich resource has no verified owner and is not scored).}
\label{tab:breakdown}
\begin{tabular}{@{}lrrr@{}}
\toprule
 & Resources & Architecture & Baseline agent \\
\midrule
Data and storage & 21 & 95\% & 44\% \\
Compute and containers & 20 & 94\% & 59\% \\
Security and secrets & 14 & 93\% & 66\% \\
Network foundation & 18 & 89\% & 45\% \\
Events, messaging and operations & 13 & 89\% & 72\% \\
Traffic and APIs & 14 & 88\% & 61\% \\
\midrule
Rich & 60 & 90\% & 54\% \\
Thin & 17 & 96\% & 66\% \\
Metadata only or opaque & 23 & 94\% & 50\% \\
\bottomrule
\end{tabular}
\end{table}

\section{Discussion}
\label{sec:discussion}

Each challenge posed in the introduction maps to a result. \emph{Scattered signal (C1):} the owners the architecture adds are found in systems other than the resource itself; code history backs 46\% of its verified owners against 2\%, and person directories 68\% against 17\%. \emph{Organization-specific conventions (C2):} recall does not fall where no signal procedure applies (94\% against 90\% on rich resources); whether profiling transfers to a second organization is untested. \emph{Multi-hop links (C3):} of the 88 direct workers that only the architecture found, 68 rest on evidence citing the resource's own stack, service or component descriptors, which context inference turns into search terms, and 42 on a ticket together with its pull request or commit. \emph{Relation strength (C4):} under the pattern weights, the advantage lies in direct work (131 of 141 people against 53), while few proposals of either system are rejected (12 and 10).

The gap is largest where ownership is not written down. On resources with an explicit owner, both systems find the named person; on resources without one, the architecture finds 117 of 123 direct workers against 47. Derived ownership adds most where declared ownership is absent.

The evaluation has limits that its design bounds. It covers one organization, with one cloud provider and 100 resources. The architecture learns an organization's ownership conventions by profiling its data rather than from rules written for it, but how well profiling works on other organizations is untested. No source is a CMDB, so derived ownership is not compared with declared ownership. Recall is measured against the owners that the two systems found together, so owners neither system found are not counted and true recall may be lower than reported; the comparison between the two systems remains reliable, as it does for pooled evaluation in information retrieval \citep{zobel1998reliable}.

Future work extends the evaluation along three axes: more organizations, to test how well profiling learns conventions it has not seen; more cloud providers, Microsoft Azure and Google Cloud, whose resource models and audit records differ from those of AWS; and more data sources, including CMDBs, service catalogs, on-call schedules and monitoring systems such as Datadog, so that derived ownership can be compared with declared ownership and drawn from a wider set of records. Each axis tests the same claim: that ownership can be derived from the records an organization already keeps.

\section{Conclusion}
\label{sec:conclusion}

Ownership is the organizational context that decides which people a security finding reaches: those responsible for fixing it and accountable for the resource it concerns. Declared ownership ages, and in the evaluated organization most ownership was implicit, held in work records, code history, audit logs and identity directories rather than on the resource.

This paper made two contributions. First, it defined ownership as a set of entities, people and teams, typed by five relation patterns ordered by strength. Second, it presented a reference architecture for agentic ownership attribution: offline, it profiles an organization into tested signal procedures and an ownership playbook; online, an investigation agent with parallel investigators attributes each resource, while code checks its conclusions and computes its confidence.

On 100 resources of 50 AWS types, the architecture reaches 92.1\% weighted recall and 86.6\% weighted precision against a pool shared with an unprimed agent, which reaches 54.8\% and 77.4\%, and it delivers coverage at 0.55 times that agent's cost. The evidence places the difference in people who build and change resources without an explicit owner, found through work records and code history. Whether it holds in other organizations is the next question.

\bibliographystyle{unsrtnat}
\bibliography{references}

\end{document}